\documentclass[prc,aps,twocolumn,showpacs,amssymb,superscriptaddress,float]{revtex4}

\usepackage{amsmath}
\usepackage{graphicx}
\usepackage{dcolumn}
\usepackage{float}
\begin{document}

\title{Reduced regulator dependence of neutron-matter predictions with
  perturbative chiral interactions}

\author{L. Coraggio}
\affiliation{Istituto Nazionale di Fisica Nucleare, \\
Complesso Universitario di Monte  S. Angelo, Via Cintia - I-80126 Napoli,
Italy}
\author{J. W. Holt}
\affiliation{Physik Department, Technische Universit\"at M\"unchen \\
D-85747 Garching, Germany}
\affiliation{Physics Department, University of Washington \\
Seattle, Washington 98195, USA}
\author{N. Itaco}
\affiliation{Istituto Nazionale di Fisica Nucleare, \\
Complesso Universitario di Monte  S. Angelo, Via Cintia - I-80126 Napoli,
Italy}
\affiliation{Dipartimento di Scienze Fisiche, Universit\`a
di Napoli Federico II, \\
Complesso Universitario di Monte  S. Angelo, Via Cintia - I-80126 Napoli,
Italy}
\author{R. Machleidt}
\affiliation{Department of Physics, University of Idaho\\
Moscow, Idaho 83844, USA}
\author{F. Sammarruca}
\affiliation{Department of Physics, University of Idaho\\
Moscow, Idaho 83844, USA}

\date{\today}

\begin{abstract}
We calculate the energy per particle in infinite neutron matter
perturbatively using chiral N$^3$LO two-body potentials plus N$^2$LO
three-body forces.
The cutoff dependence of the predictions is investigated by employing
chiral interactions with different regulators.
We find that the inclusion of three-nucleon forces, which are
consistent with the applied two-nucleon interaction, leads to a
strongly reduced  regulator dependence of the results.
\end{abstract}


\maketitle

\section{Introduction}
A major breakthrough in the last decade has been the derivation of
nucleon-nucleon ($NN$) potentials, $V_{NN}$, based on chiral
perturbation theory (ChPT) that are able to reproduce accurately the
$NN$ data \cite{EM03,EGM05,ME11}.

The idea to construct realistic two- and three-nucleon forces (2NF and
3NF) starting from a chiral Lagrangian goes back to the seminal
work of Weinberg \cite{Wei79,Wei90,Wei91}, who invoked the concept of
an effective field theory (EFT) to study the $S$-matrix for processes
involving arbitrary numbers of low-momentum pions and nucleons.
In this approach, the long-range forces are ruled by the symmetries of
low-energy QCD (particularly, spontaneously broken chiral symmetry),
and the short-range dynamics is absorbed into a complete basis of
contact terms that are proportional to low-energy constants (LECs) fit
to 2N data.
 
One great advantage of ChPT is that it generates nuclear two- and
many-body forces on an equal footing~\cite{Wei92,Kol94,ME11}.
Most interaction vertices that appear in the 3NF and in the
four-nucleon force (4NF) also occur in the 2NF.
The parameters carried by these vertices are fixed (along with the
LECs of the 2N contact terms) in the construction of the chiral 2NF.
Consistency then requires that for the same vertices the same
parameter values are used in the 2NF, 3NF, 4NF, \ldots .

A crucial theme in EFT is regulator independence within the range of
validity of the theory.
In other words, the physical observables calculated in the theory must
be independent both of the choice of the regulator function as well as
its cutoff scale $\Lambda$.
ChPT is a low-momentum expansion which is valid only for momenta $ Q <
\Lambda_\chi \simeq 1$ GeV, where $\Lambda_\chi$ denotes the chiral
symmetry breaking scale.
Therefore, $NN$ potentials derived in this framework are usually
multiplied by a regulator function
\begin{equation}
f(p',p) = \exp [-(p'/\Lambda)^{2n} - (p/\Lambda)^{2n}] \,,
\label{eq_reg}
\end{equation} 
where typical choices for the cutoff parameter are $\Lambda \simeq
0.5$ GeV.
In regards to the physics of the two-nucleon problem, it is obvious
that the solutions of the Lippmann-Schwinger equation, that are
related to the two-nucleon observables, may depend sensitively on the
regulator and its cutoff parameter.
This unwanted dependence is then removed by a renormalization
procedure, in which the contact terms are re-adjusted to reproduce the
two-nucleon phase shifts and data.
However, it is well-known that phase equivalent potentials do not
necessarily yield identical results in the many-body problem. 
Thus, one may be confronted with cutoff dependence in the many-body
system \cite{Cor12}.
However, in the many-body problem, also 3NF, 4NF, ... contribute,
which will have impact on the final predictions and may either
increase or reduce the cutoff dependence.

A convenient theoretical laboratory to investigate this issue is
infinite nuclear matter and neutron matter.
The advantage of pure neutron matter is that the contact interaction,
$V_E$, and the  1$\pi$-exchange term, $V_D$, that appear in the
N$^2$LO three-body force, vanish~\cite{HS10}. 
Thus, the low-energy constants of $V_E$ and $V_D$ (known as $c_E$ and
$c_D$), which cannot be constrained by two-body observables, are not
needed.
Consequently, the calculation of the ground state energy of infinite
neutron matter, with chiral 3NFs up to N$^2$LO, depends only on
parameters that have been fixed in the two-nucleon system.

We note that there have been already some attempts to study the
uncertainties in neutron matter predictions using chiral forces,
e.~g., by Hebeler and Schwenk~\cite{HS10}, and Tews {\it et al.}
\cite{Tews12}, who come up with uncomfortably large
uncertainties---for reasons to be discussed below.
It is also worth noting that, aside from the above considerations,
neutron matter, and more generally isospin-asymmetric nuclear matter,
is currently of great interest in the nuclear physics community
because of its close connection with the physics of neutron-rich
nuclei and, for higher densities, with the structure of neutron stars. 

It is the purpose of the present paper, to investigate how the
equation of state of neutron matter, calculated using chiral nuclear
potentials, depends on the choice of the regulator function.
More precisely, we employ three different chiral potentials, whose
cutoff parameters are $\Lambda = 414$ \cite{Cor07}, 450, and 500 MeV
\cite{EM03,ME11} and calculate, including 3NF effects, the energy per
nucleon for neutron matter at nuclear densities in the framework of
many-body perturbation theory.
The crucial point of our calculations is that we use in the 3NF
exactly the same LECs as well as the same cutoff parameters as in the
2NF.
We will show that this consistent use of the LECs in the 2NF {\it and}
3NF leads to a substantial reduction of the regulator dependence of
the neutron matter predictions.

The paper is organized as follows. 
In Sec.~II, we briefly describe the features of the different chiral
potentials employed and, in Sec.~III, we give an outline of the
calculation of the energy per nucleon in neutron matter that takes into
account 3NF effects.
Our results are presented in Sec.~IV and some concluding remarks and
an outlook are given in Sec.~V.

\section{The chiral potentials}
During the past two decades, it has been demonstrated that chiral
effective field theory (chiral EFT) represents a powerful tool to deal
with hadronic interactions at low energy in a systematic and
model-independent way (see Refs.~\cite{ME11,EHM09} for recent reviews). 
For the construction of an EFT, it is crucial to identify a separation
of scales. 
In the hadron spectrum, a large gap between the masses of the pions
and the masses of the vector mesons, like $\rho(770)$ and
$\omega(782)$, can clearly be identified. 
Thus, it is natural to assume that the pion mass sets the soft scale,
$Q \sim m_\pi$, and the rho mass the hard scale, $\Lambda_\chi \sim
m_\rho \sim 1$ GeV, also known as the chiral-symmetry breaking scale.
This is suggestive of considering a low-energy expansion arranged in
terms of the soft scale over the hard scale, $(Q/\Lambda_\chi)^\nu$,
where $Q$ is generic for an external momentum (nucleon three-momentum
or pion four-momentum) or a pion mass.
The appropriate degrees of freedom are, obviously,  pions and
nucleons, and not quarks and gluons.
For this EFT to rise above the level of phenomenology, it must
have a firm link with QCD.
The link is established by having the EFT observe all relevant
symmetries of the underlying theory, in particular, the broken chiral
symmetry of low-energy QCD~\cite{Wei79}.
The past 15 years have seen great progress in applying ChPT to nuclear
forces.
As a result, $NN$ potentials of high precision have been constructed,
which are based upon ChPT carried to N$^3$LO.

\begin{figure}[H]
\begin{center}
\includegraphics[scale=0.35,angle=0]{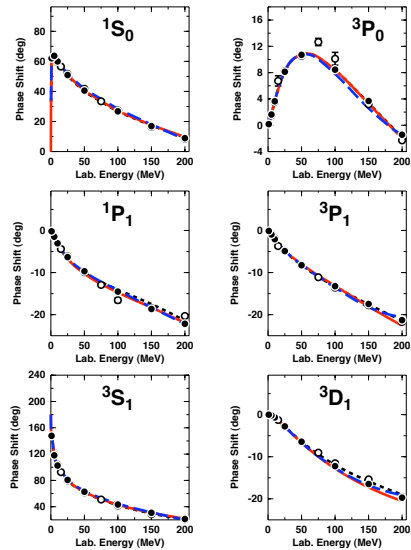}
\vspace*{-1.5cm}
\caption{(Color online) 
Neutron-proton phase parameters as predicted by chiral N$^3$LO
potentials with different cutoff scale $\Lambda$. Solid (red)
curve, $\Lambda=414$ MeV; dashed (blue) curve, $\Lambda=450$ MeV; and
dotted (black) curve, $\Lambda=500$ MeV. Partial waves with total
angular momentum $J\leq 1$ are displayed. The solid dots and open
circles are the results from the Nijmegen multi-energy $np$ phase
shift analysis~\protect\cite{Sto93}  and the VPI/GWU single-energy
$np$ analysis SM99~\protect\cite{SM99}, respectively.}
\label{ph1}
\end{center}
\end{figure}

\begin{figure}[H]
\begin{center}
\includegraphics[scale=0.35,angle=0]{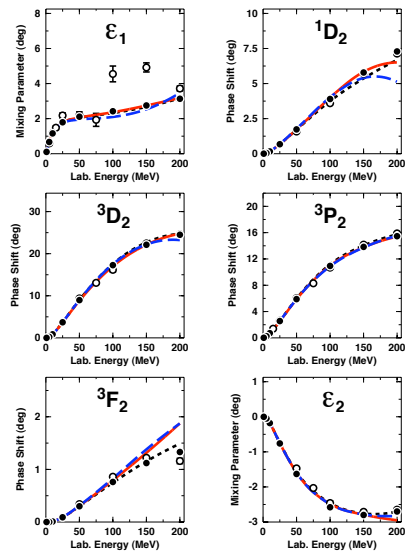}
\vspace*{-1.5cm}
\caption{(Color online) 
Same as Fig.~\ref{ph1}, but $J=2$ phase shifts and $J\leq 2$ mixing
parameters are shown.}
\label{ph2}
\end{center}
\end{figure}

Since ChPT is a low-momentum expansion, valid only for momenta $Q <
\Lambda_\chi$, the potentials are either abruptly set to zero for
momenta above a certain cutoff $\Lambda < \Lambda_\chi$ (``sharp
cutoff'') or they are multiplied with a smooth regulator function,
like, e.~g., the one of Gaussian shape given in Eq.~(\ref{eq_reg}).

\begin{figure}[H]
\begin{center}
\includegraphics[scale=0.35,angle=0]{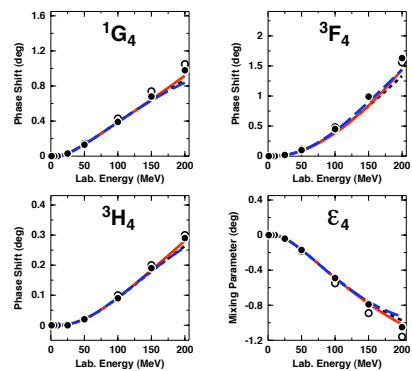}
\vspace*{-4cm}
\caption{(Color online) 
Same as Fig.~\ref{ph1}, but some representative peripheral partial
waves are shown.}
\label{ph3}
\end{center}
\end{figure}

In this investigation, we consider three N$^3$LO potentials which
differ by the cutoff parameter $\Lambda$ and/or the regulator function:
\begin{itemize}
\item
$\Lambda = 414$ MeV together with a sharp cutoff (published in
Ref.~\cite{Cor07}).
\item
$\Lambda = 450$ MeV using the regulator function Eq.~(\ref{eq_reg})
with $n=3$.
We have constructed this potential for the present investigation.
\item
$\Lambda = 500$ MeV using the regulator function Eq.~(\ref{eq_reg})
with $n=2$ for the 2$\pi$ exchange contributions. 
This potential was published in 2003~\cite{EM03}.
\end{itemize}
All three potentials use the same (comprehensive) analytic expressions
which can be found in Ref.~\cite{ME11}.
Note that the Gaussian regulator function Eq.~(\ref{eq_reg})
suppresses the potential also for $Q<\Lambda$, which is why we use a
sharp cutoff function in the case of the lowest cutoff of 414 MeV.
Cutoff-independence is an important aspect of an EFT.
In lower partial waves, the cutoff dependence of the $NN$ phase shifts
is counter balanced by an appropriate adjustment of the contact terms
which, at N$^3$LO, contribute in $S$, $P$, and $D$ waves. 
The extent to which cutoff independence can be achieved in lower
partial waves is demonstrated in Figs.~\ref{ph1}-\ref{ph2}.
In $F$ and higher partial waves (where there are no $NN$ contact
terms) the LECs of the dimension-two $\pi N$ Lagrangian can be used to
obtain cutoff independence of the phase shift predictions, see
Table~\ref{tab1} and Fig.~\ref{ph3}.

\begin{table}[H]
\caption{For the various chiral N$^3$LO $NN$ potentials used in the
  present investigation, we show the cutoff parameter $\Lambda$, the
  type of regulator, the exponent $n$ used in the regulator function,
  Eq.~(\ref{eq_reg}), and the LECs  of the dimension-two $\pi N$
  Lagrangian, $c_i$ (in units of GeV$^{-1}$), which are relevant for
  the N$^2$LO 3NF in neutron matter.}
\label{tab1}
\smallskip
\begin{ruledtabular}
\begin{tabular}{cccc}
\noalign{\smallskip}
               & \multicolumn{3}{c}{Cutoff parameter $\Lambda$ (MeV)} \\
               \cline{2-4}
                 &      414                     & 450           &    500            \\
\hline
\noalign{\smallskip}
Regulator type & sharp & Gaussian & Gaussian \\
n & -- & 3 & 2 \\
$c_1$ & --0.81 & --0.81 & -0.81 \\
$c_3$ & --3.00 & --3.40 &  --3.20 \\
\noalign{\smallskip}
\end{tabular}
\end{ruledtabular}
\end{table}

An important advantage of the EFT approach to nuclear forces is that
it creates two- and many-body forces on an equal footing.
The first non-vanishing 3NF occurs at N$^2$LO. 
At this order, there are three 3NF topologies: the two-pion exchange
(2PE), one-pion exchange (1PE), and 3N-contact interactions. 
The 2PE 3N-potential is given by
\begin{equation}
V_{\rm c} = 
\left( \frac{g_A}{2f_\pi} \right)^2
\frac12 
\sum_{i \neq j \neq k}
\frac{
( \vec \sigma_i \cdot \vec q_i ) 
( \vec \sigma_j \cdot \vec q_j ) }{
( q^2_i + m^2_\pi )
( q^2_j + m^2_\pi ) } \;
F^{ab}_{ijk} \;
\tau^a_i \tau^b_j
\label{eq_3nf_nnloa}
\end{equation}
with $\vec q_i \equiv \vec{p_i}' - \vec p_i$, where $\vec p_i$ and
$\vec{p_i}'$ are the initial and final momenta of nucleon $i$,
respectively, and 
\begin{equation}
F^{ab}_{ijk} = \delta^{ab}
\left[ - \frac{4c_1 m^2_\pi}{f^2_\pi}
+ \frac{2c_3}{f^2_\pi} \; \vec q_i \cdot \vec q_j \right]
+ 
\frac{c_4}{f^2_\pi}  
\sum_{c} 
\epsilon^{abc} \;
\tau^c_k \; \vec \sigma_k \cdot [ \vec q_i \times \vec q_j] \; .
\label{eq_3nf_nnlob}
\end{equation}  
Note that the 2PE 3NF does not contain any new parameters, because the
LECs $c_1$, $c_3$, and $c_4$ appear already  in the 2PE 2NF.
The 1PE contribution is
\begin{equation}
V_{D} = 
-\frac{c_D}{f^2_\pi\Lambda_\chi} \; \frac{g_A}{8f^2_\pi} 
\sum_{i \neq j \neq k}
\frac{\vec \sigma_j \cdot \vec q_j}{
 q^2_j + m^2_\pi }
( \mbox{\boldmath $\tau$}_i \cdot \mbox{\boldmath $\tau$}_j ) 
( \vec \sigma_i \cdot \vec q_j ) 
\label{eq_3nf_nnloc}
\end{equation}
and the 3N contact potential reads
\begin{equation}
V_{E} = 
\frac{c_E}{f^4_\pi\Lambda_\chi}
\; \frac12
\sum_{j \neq k} 
 \mbox{\boldmath $\tau$}_j \cdot \mbox{\boldmath $\tau$}_k  \; .
\label{eq_3nf_nnlod}
\end{equation}
In the above, we use $g_A=1.29$, $f_\pi=92.4$ MeV, $m_\pi=138.04$ MeV,
and $\Lambda_\chi=700$ MeV.
The last two 3NF terms involve the two new parameters $c_D$ and $c_E$,
which do not appear in the 2N problem.
There are many ways to pin these two parameters down.
The triton binding energy and the $nd$ doublet scattering length
$^2a_{nd}$ can be used.
Alternatively, one may choose the binding energies of $^3$H and $^4$He
or an optimal over-all fit of the properties of light nuclei.
However, in neutron matter, $V_D$ and $V_E$ do not contribute such
that we do not have to worry about their values here.
Note also that the $c_4$ term of $V_c$, Eqs.(\ref{eq_3nf_nnloa}) and
(\ref{eq_3nf_nnlob}), vanishes in neutron matter.

\section{Calculation of the energy per particle in neutron matter}
We calculate the ground-state energy per particle (g.s.e.) of infinite
neutron matter within the framework of many-body perturbation theory.
In particular, we express the g.s.e. as a sum of Goldstone diagrams up
to third order.

\begin{figure}[H]
\begin{center}
\includegraphics[scale=0.50,angle=-90]{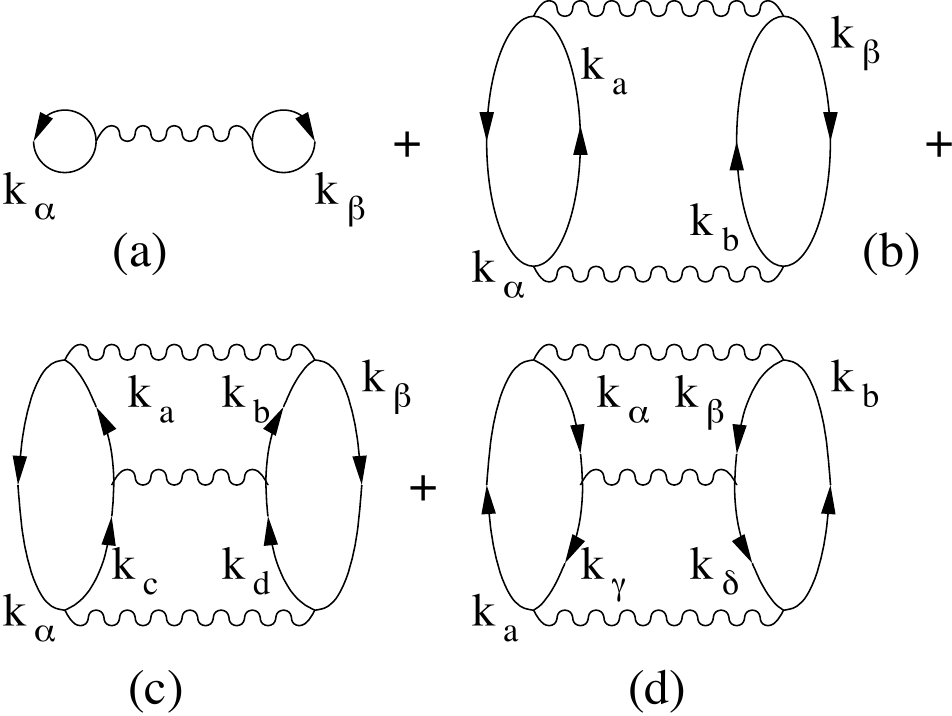}
\caption{First-, second-, and third-order diagrams of the Goldstone
  expansion included in our calculations with $V_{NN}$ vertices
  only. Latin-letter subscripts denote particle states, greek-letter
  subscripts correspond to hole states.}
\label{figgold2}
\end{center}
\end{figure}

In order to take into account the effects of the N$^2$LO 3NF, a
density-dependent two-body potential ${\overline V}_{NNN}$ is added to
the chiral N$^3$LO potential $V_{NN}$.
This potential ${\overline V}_{NNN}$ is obtained by summing one
nucleon over the filled Fermi sea, which leads to a density-dependent
two-nucleon interaction~\cite{HKW09,HKW10}.
Hebeler {\it et al.}~\cite{HS10} have pointed out that to take care of
the correct combinatorial factors of the normal-ordering at the
two-body level of the 3NF, the matrix elements of ${\overline
  V}_{NNN}(k_F)$ are to be multiplied by a factor 1/3 in the
first-order Hartree-Fock (HF) diagram, and by a factor 1/2  in the
calculation of the single-particle energies (s.p.e.).

\begin{figure}[H]
\vspace{-3.0truecm}
\hspace{-2.0truecm}
\includegraphics[scale=0.45,angle=0]{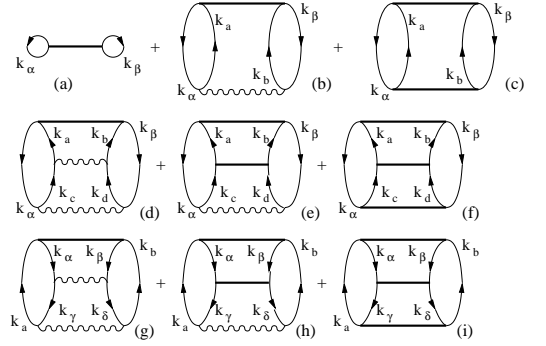}
\vspace{-3.0truecm}
\caption{Same as in Fig. \ref{figgold2}, but also including
  $\overline{V}_{NNN}$ vertices which are denoted with thick straight
  lines.}
\label{figgold3}
\end{figure}

In Fig. \ref{figgold2} we show the diagrams we have included in our
calculation, where only the $V_{NN}$ vertices are taken into account.
The only diagram we do not include is the third-order $ph$ diagram.
The diagrams that include the effects of $V_{NNN}$ are shown in
Fig. \ref{figgold3}.

The first-order HF contribution is explicitly given by

\begin{eqnarray}
E_1 &=& \frac{8}{\pi} \int_0^{k_F} k^2 dk \left[ 1 - \frac{3}{2}
  \frac{k}{k_F} + \frac{1}{2} \left( \frac{k}{k_F} \right)^3 \right]
\sum_{JLS} \nonumber \\
~&~&(2J+1) [V^{JLLS}_{NN}(k,k)+ \frac{1}{3}
{\overline V}^{JLLS}_{NNN}(k,k)]~~.~~~~~~~~~~~
\end{eqnarray}
\label{HF}

The second-order diagrams are computed using the so-called
angle-average (AA) approximation \cite{Lom63}, and their contribution
is

\begin{eqnarray}
E_2 & = & - \frac{6}{\pi^2k_F^3} \int_0^{2k_F} K^2dK \int_0^{\infty}
k'^2dk' \int_0^{\infty} k^2dk \nonumber \\
~&~&P(k',K) Q(k,K)  \sum_{JL{\overline L}S} (2J+1) \nonumber \\ 
~&~&\frac{[V^{JL{\overline L}S}_{NN}(k,k')+
  {\overline V}^{JL{\overline L}S}_{NNN}(k,k')]^2}{E(k,k',K)} ~~.
\label{2ndorder}
\end{eqnarray}

The operators $P$ and $Q$ are defined through the relationships:

\begin{eqnarray}
Q(k,K) = & 0~~, & 0 \leq k \leq (k_F^2 - \frac{K^2}{4})^{1/2} \nonumber \\
 ~~    = & -\frac{k_F^2-k^2-K^2/4}{kK}~~, & (k_F^2 -
 \frac{K^2}{4})^{1/2} \leq k \leq (k_F + \frac{K}{2}) \nonumber \\
 ~~    = & 1~~, & k \geq (k_F + \frac{K}{2}) \nonumber \\
P(k,K) = & 1~~, & 0 \leq k \leq (k_F - \frac{K}{2}) \nonumber \\
 ~~    = & \frac{k_F^2-k^2-K^2/4}{kK}~~, & (k_F - \frac{K}{2}) \leq
 k \leq (k_F^2 - \frac{K^2}{4})^{1/2} \nonumber \\
~~     = & 0~~, & k \geq (k_F^2 - \frac{K^2}{4})^{1/2} \,.  \nonumber
\end{eqnarray}

\noindent
In Eq.~\ref{2ndorder}, the denominator is $E(k,k',K) =
\frac{\hbar^2k'^2}{M}+2U\left(\sqrt{\frac{K^2}{4}+k'^2}\right) -
\frac{\hbar^2k^2}{M}-2U \left(\sqrt{\frac{K^2}{4}+k^2}\right)$,
$U(\tilde{k})$ being the self-consistent single-particle potential:

\begin{eqnarray}
U(\tilde{k}) & = & 8 \sum_{JLLS} (2J+1)^2 \left\{
  \left[ \int_0^{\frac{1}{2}(k_F-\tilde{k})} \tilde{k}'^2 d\tilde{k}'
    + \right. \right. \nonumber \\
~&~& \left.\left. \frac{1}{2\tilde{k}}
    \int_{\frac{1}{2}(k_F-\tilde{k})}^{\frac{1}{2}(k_F+\tilde{k})}
    \tilde{k}'d\tilde{k}' (\frac{1}{4} (k_F^2 -\tilde{k}^2) -
    \tilde{k}'(\tilde{k}'-\tilde{k})) \right] \right.
\nonumber \\
~&~& \left. \left[ V^{JLLS}_{NN}(\tilde{k}',\tilde{k}')+ \frac{1}{2}
    {\overline V}^{JLLS}_{NNN}(\tilde{k}',\tilde{k}') \right]
\right\} ~~.
\end{eqnarray}

The particle-particle ($pp$) and hole-hole ($hh$) third-order diagrams
are also computed in the AA approximation, and their explicit
expressions are:

\begin{eqnarray}
E_3(pp) & = & \frac{12}{(\pi k_F)^3} \int_0^{2k_F} K^2dK \int_0^{\infty}
k^2dk \int_0^{\infty} k'^2dk' \nonumber \\ 
~&~& \int_0^{\infty} k''^2dk'' P(k,K) Q(k',K) Q(k'',K)
\nonumber \\
~&~& \sum_{JL{\overline L}{\overline L'}S} (2J+1) [V^{JL{\overline L}S}_{NN}(k,k') + \nonumber \\ 
~&~& {\overline V}^{JL{\overline L}S}_{NNN}(k,k')] [V^{J{\overline L}{\overline L'}S}_{NN}(k',k'') + \nonumber \\ 
~&~& {\overline
    V}^{J{\overline L}{\overline  L'}S}_{NNN}(k',k'')] [V^{J{\overline L'}LS}_{NN}(k'',k)  \nonumber \\
~&~& +{\overline V}^{J{\overline
      L'}LS}_{NNN}(k'',k)]/ [E(k'',k) \cdot E(k',k)] ~~,
\end{eqnarray}
\label{3rdorderpp}

\begin{eqnarray}
E_3(hh) & = & \frac{2}{(\pi k_F)^3} \int_0^{2k_F} K^2dK \int_0^{\infty}
k^2dk \int_0^{\infty} k'^2dk' \nonumber \\
~&~&\int_0^{\infty} k''^2dk'' P(k,K) Q(k',K) P(k'',K) \nonumber \\
~&~& \sum_{JL{\overline L}{\overline L'}S} (2J+1) [V^{JL{\overline L}S}_{NN}(k,k') +  \nonumber \\
~&~& {\overline V}^{JL{\overline L}S}_{NNN}(k,k')] [V^{J{\overline L}{\overline L'}S}_{NN}(k',k'') +  \nonumber \\
~&~&{\overline
    V}^{J{\overline L}{\overline L'}S}_{NNN}(k',k'')][V^{J{\overline L'}LS}_{NN}(k'',k) + \nonumber \\
~&~&{\overline V}^{J{\overline
    L'}LS}_{NNN}(k'',k)] / [E(k',k'') \cdot E(k',k)] ~~.
\end{eqnarray}
\label{3rdorderhh}

We have also calculated the $[2|1]$ Pad\'e approximant \cite{BG70} 

\begin{equation}
E_{[ 2|1 ]}=E_0+ E_1+\frac{E_2}{1-E_{3}/E_2}~~,
\end{equation}

\noindent
$E_i$ being the $i$th order energy contribution in the perturbative
expansion of the g.s.e..
The Pad\'e approximant is an estimate of the value to which the
perturbative series may converge.
Thus,  the comparison between the third-order results and those
obtained by means of the $[2|1]$ Pad\'e approximant provides an
indication of the size of the higher-order perturbative terms.
It is worth mentioning that the role of Pad\'e approximants in
many-body perturbation theory for nuclear systems has been explored in
the last decade for finite nuclei \cite{MSTA1,MSTA2,Roth1,Roth2}.

\section{Results}
As explained in the previous section, we calculate the energy per
particle of neutron matter in the framework of many-body perturbation
theory, including contributions up to third-order in the interaction.
Therefore, it is of interest to obtain an idea of the convergence 
of the perturbative expansion of the g.s.e..

\begin{figure}[h]
\begin{center}
\vspace{-1.5truecm}
\includegraphics[scale=0.40,angle=0]{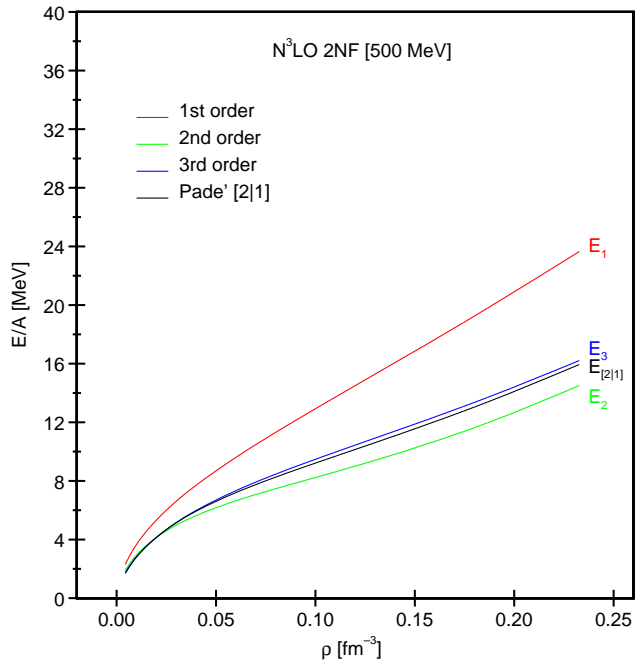}
\caption{(Color online) Neutron matter energy per particle obtained
  from the N$^3$LO 2NF with cutoff $\Lambda=500$ MeV. The first,
  second, and third order in the perturbative expansion and the Pad\'e
  approximant $[2|1]$ are shown as a function of density $\rho$.}
\label{conv1}
\end{center}
\end{figure}

In Fig.~\ref{conv1}, we show the neutron-matter energy per nucleon as
a function of density, calculated at various orders in the perturbative
expansion applying the chiral N$^3$LO $NN$ potential with a cutoff
parameter equal to 500 MeV.
We have chosen here the potential with the largest cutoff since it has
the worst perturbative behavior.
From the inspection of Fig. \ref{conv1}, it can be seen that the energy
per nucleon calculated at second order, $E_2$, does not differ much
from the one computed at third order, $E_3$, for the whole range of
densities shown.
The perturbative character is also indicated by the fact that $E_3$ is
quite close to the energy obtained with the $[2|1]$ Pad\'e approximant.

For completeness, we mention that we also performed calculations
employing the chiral N$^3$LO $NN$ potential with a cutoff parameter
equal to 600 MeV \cite{ME11}, but we found its perturbative behavior
unsatisfactory, in agreement with the observations by Tews et
al. \cite{Tews12}.

\begin{figure}[h]
\begin{center}
\vspace{-1.5truecm}
\includegraphics[scale=0.40,angle=0]{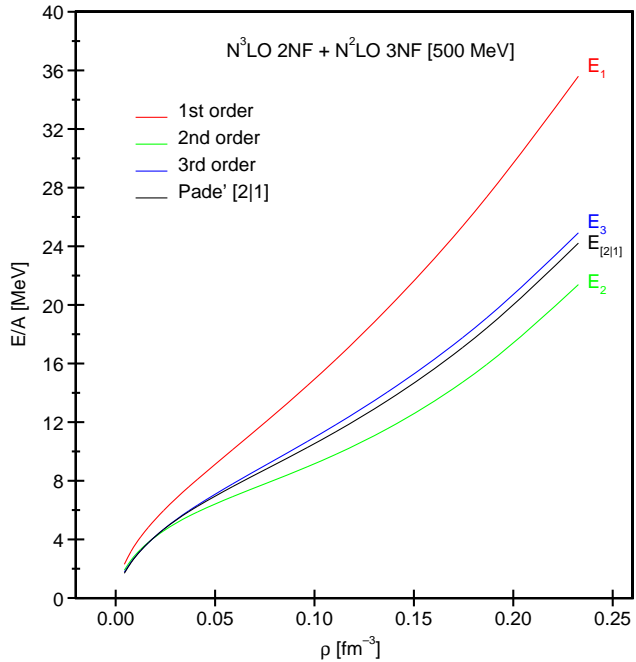}
\caption{(Color online) Same as in Fig. \ref{conv1}, but including the
  contribution of the N$^2$LO 3NF.}
\label{conv2}
\end{center}
\end{figure}

We have also investigated the perturbative behavior of our
calculations when including the effects of $V_{NNN}$.
Fig. \ref{conv2} shows that, starting from the same N$^3$LO potential,
there is a small enhancement of the higher-order terms when including
the N$^2$LO 3NF.
Nevertheless, the results at third order are very close to those
obtained with the $[2|1]$ Pad\'e approximant.

Our main goal is to calculate the g.s.e. per particle in infinite
neutron matter, starting from N$^3$LO chiral $NN$ potentials that
apply different regulator functions.
This is done by using the chiral potentials introduced in Sec.~II.
We have added to each 2NF a chiral N$^2$LO 3NF whose low-energy
constants $c_1$ and $c_3$, cutoff parameters, and regulator function
are exactly the same as in the corresponding N$^3$LO $NN$ potential,
see Table~\ref{tab1}.

\begin{figure}[h]
\begin{center}
\vspace{-1.5truecm}
\includegraphics[scale=0.40,angle=0]{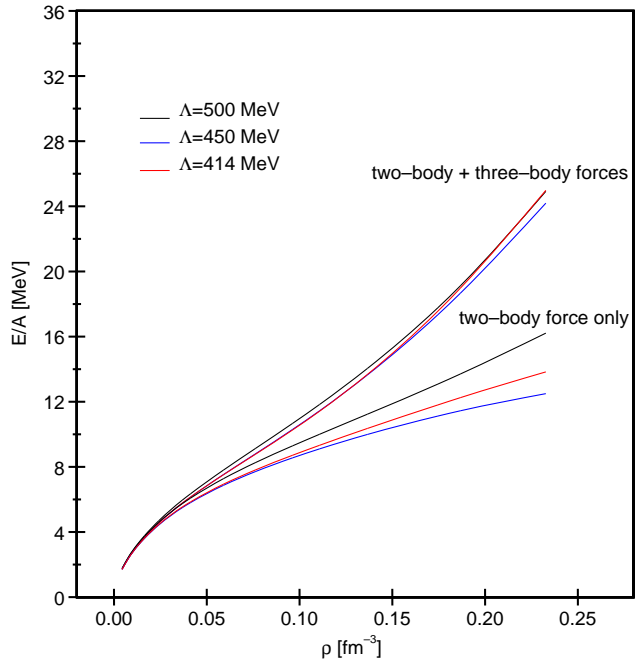}
\caption{(Color online) Results obtained for the g.s.e. per particle
  of infinite neutron matter at third-order in perturbation theory
  for three sets of chiral interactions which differ by the cutoff
  parameter $\Lambda$.}
\label{comparison}
\end{center}
\end{figure}

In Fig. \ref{comparison}, we show our results, obtained at third-order
in the perturbative expansion, with and without taking into account
3NF effects.
The results obtained with 2NFs show considerable dependence on the
choice of the regulator and its cutoff parameter.
This is at variance with the desired regulator independence of the
EFT.
However, when including the contributions of the three-body potentials,
which are consistent with their 2NF partner, regulator dependence is
strongly reduced.
This is our main result and, at the same time,
the first clear evidence that modern chiral potentials can
provide model-independent results in many-body calculations
{\it if 2NF and 3NF are treated consistently.}

\section{Concluding remarks and outlook.}
In this paper we have studied the regulator dependence of many-body
predictions when employing chiral two- and three-nucleon potentials,
using as a testing ground the perturbative calculation of the
neutron-matter energy per particle.
We find substantial regulator dependence of the predictions when only
2NFs are taken into account. 
The main outcome of this study is the observation that the 3NF can
play a crucial role in the restoration of regulator independence. 
However, this mechanism works properly only when the chiral 2NF and
3NF are treated consistently in the sense that the same parameters are
used for the same vertices that occur in all topologies involved. 
This is particularly true for the LECs $c_1$ and $c_3$ occurring first
at N$^2$LO in the chiral power counting.

In Refs. \cite{HS10,Tews12} the large uncertainties of the results for
the ground-state energy per neutron trace back to the choice of using
a range of values for $c_1$ and $c_3$ obtained from a high-order
analysis of $\pi N$ scattering \cite{Krebs12}. 
This is at variance with the $c_i$s employed in the present paper which,
as reported in Section II, are uniquely fixed in peripheral $NN$
partial waves.

In closing, we note that the present investigation deals only with
identical nucleon systems, and that the regulator dependence should
also be investigated in systems with different concentrations of
interacting protons and neutrons. 
In infinite symmetric nuclear matter also contributions from the
intermediate-range 1$\pi$-exchange component $V_D$, and from the
short-range contact interaction $V_E$ come into play.
This means that the calculation of the g.s.e. depends also on the
coupling constants $c_D$ and $c_E$.
Even though these parameters can be fixed in few-body systems, there
is some freedom in doing so, resulting in more latitude for the 3NF
contribution in nuclear matter (as compared to pure neutron matter). 

This will be an interesting subject for a future study, that may shed
more light on the topic of regulator independence of many-body
calculations with chiral potentials. 
The results of such investigations will provide valuable guidance for
the proper application of these interactions in microscopic nuclear
structure calculations.

\subsection*{Acknowledgments}
This work was supported in part by the U.S. Department of Energy under
Grant No.\ DE-FG02-03ER41270 and No.\ DE-FG02-97ER-41014, by the
Italian Ministero dell'Istruzione, dell'Universit\`a e della Ricerca
(MIUR) under PRIN 2009, by BMBF (the DFG cluster of excellence: Origin
and Structure of the Universe), and by DFG and NSFC (CRC110). 


\end{document}